\documentclass[twocolumn,a4paper,10pt]{article}
\usepackage{amsmath,amsfonts}
\usepackage{epsfig}
\usepackage{times}
\usepackage{epsfig}
\usepackage[small,hang]{caption2}
\usepackage{graphicx}
\usepackage[normalem]{ulem}
\usepackage[dvipsnames]{xcolor}
\definecolor{brown}{rgb}{0.5,0.1,0.05}

\makeatletter

\newcounter{numbersec}
\renewcommand{\section}[1]{\par\noindent\stepcounter{numbersec}
\par
\vspace{6pt}
\noindent\textbf{\large   \arabic{numbersec} \hspace*{0.3cm} #1 }
\par
\vspace{2pt}
}
\renewcommand{\subsection}[1]{
\par
\vspace{6pt}
\noindent\textbf{#1}
\par
}
\renewcommand{\subsubsection}[1]{%
\par
\vspace{6pt}
\textbf{#1.}
}

\newcommand{\Abstract}{\par\vspace{6pt}\noindent\textbf{\large Abstract}\par\vspace{2pt}}
\newcommand{\Acknowledgments}{\par\vspace{6pt}\noindent\textbf{\large Acknowledgments }\par\vspace{2pt}}

\newenvironment{References}{
\par\vspace{6pt}\noindent\textbf{\large References}\par\vspace{2pt}
\begin{small}\begin{list}{ }{
\itemsep0mm \parsep0mm\labelsep0mm\leftmargin0mm
}}
{\end{list}\end{small}}

\makeatother

\title{\vspace*{-12mm}
\LARGE \sc \textbf{  
Predicting the near-wall region of turbulence through convolutional neural networks
}}
\author{ \Large \bf \textit{ 
A.G. Balasubramanian$^{1,2}$, L. Guastoni$^{1,2}$, A. G\"uemes$^{3}$, A. Ianiro$^{3}$,} \\ \Large \bf \textit{S. Discetti$^{3}$, P. Schlatter$^{1,2}$, H. Azizpour$^{4,2}$ and R. Vinuesa$^{1,2*}$} \\ \\
\bf  $^{1}$ \textit{SimEx/FLOW, Engineering Mechanics, KTH Royal Institute of Technology, Sweden} \\
\bf  $^{2}$ \textit{Swedish e-Science Research Centre (SeRC), Sweden} \\
\bf  $^{3}$ \textit{Aerospace Engineering Research Group, Universidad Carlos III de Madrid, Spain} \\ 
\bf $^{4}$ \textit{Division of Robotics, Perception, and Learning, KTH Royal Institute of
Technology, Sweden} \\
\underline{\bf rvinuesa@mech.kth.se}
}
\date{}

\begin{document}

\maketitle
\thispagestyle{empty}

%
%
\Abstract
Modelling the near-wall region of wall-bounded turbulent flows is a widespread practice to reduce the computational cost of large-eddy simulations (LESs) at high Reynolds number. As a first step towards a data-driven wall-model, a neural-network-based approach to predict the near-wall behaviour  in a turbulent open channel flow is investigated.
The fully-convolutional network (FCN) proposed by Guastoni {\it et al.} (2020{\it b}) is trained to predict the two-dimensional velocity-fluctuation fields at $y^{+}_{\rm target}$, using the sampled fluctuations in wall-parallel planes located farther from the wall, at $y^{+}_{\rm input}$.
The data for training and testing is obtained from a direct numerical simulation (DNS) at friction Reynolds numbers $Re_{\tau} = 180$ and $550$. The turbulent velocity-fluctuation fields are sampled at various wall-normal locations, {\it i.e.} $y^{+} = \{15, 30, 50, 80, 100, 120, 150\}$. At $Re_{\tau}=550$, the FCN can take advantage of the self-similarity in the logarithmic region of the flow and predict the velocity-fluctuation fields at $y^{+} = 50$ using the velocity-fluctuation fields at $y^{+} = 100$ as input with less than 20\% error in prediction of streamwise-fluctuations intensity. 
These results are an encouraging starting point to develop a neural-network based approach for modelling turbulence at the wall in numerical simulations.

%
%
\section{Introduction}
Turbulent flows in numerous engineering applications are characterized by a very high Reynolds number and currently, they cannot be simulated numerically without introducing some form of modelling. This is due to the computational cost arising from the resolution requirements to simulate all the length and time scales in turbulent flows. Large-eddy simulations (LESs) address this issue by filtering out the smallest turbulent scales and modelling the eddies that are smaller than the filter size.
This approach can be problematic in wall-bounded turbulent flows, where the smaller scales characterize the flow and it becomes computationally expensive to perform LESs at high Reynolds number. This is due to the resolution requirements for resolving the dynamically important flow structures in the viscous and the logarithmic layers, which scale as $\mathcal{O} (Re_{\tau}^2)$ (Larsson {\it et al.} 2015), where the friction Reynolds number is defined in terms of reference length $h$ and friction velocity $u_{\tau}$. 
Since a wall-resolved LES can be prohibitively expensive from the computational point of view, several methods have been proposed to model the near-wall region and simulate complex flows at realistic Reynolds numbers. 
These wall models aim to reproduce the most important features of the inner-layer dynamics like (\textit{e.g.} streaks and streamwise vortices), without integrating the Navier--Stokes equations in that flow region. A classification of wall models with the corresponding assumptions and error sources is presented in Piomelli and Balaras (2002).
A recent review of wall models can be found in Larsson {\it et al.} (2015).

One notable approach was proposed by Mizuno and Jim\'enez (2013): by taking advantage of the self-similarity hypothesis, according to which the eddy sizes scale linearly in the logarithmic region at high Reynolds number, an \emph{off-wall} boundary condition was defined. In their work, the velocity field at the off-wall boundary location is substituted by a re-scaled and a shifted copy of an interior reference plane farther from the wall. This allows the logarithmic region to be simulated without considering the near-wall dynamics.

In this work, a preliminary investigation is conducted to a develop neural-network model that predicts the flow close to the wall based on information farther from the wall, as a first step towards the definition of an off-wall boundary condition.
Neural networks are deep-learning models that have started offering new interesting opportunities to formulate efficient data-driven wall models, as in Beck \textit{et al.} (2019) and Moriya \textit{et al.} (2021). Once the model is trained, the evaluation of the neural network is computationally cheap and they can provide a valid alternative to the models that are currently employed within numerical simulations. 
In this study the neural-network architecture of choice is the fully-convolutional network (FCN) proposed by Guastoni {\it et al.} (2020{\it b}), which was originally designed to predict the turbulent velocity field at a given wall-normal distance, using quantities measured at the wall as inputs.
We test the performance of the FCN architecture on a task that is different, yet similar, to the one it was designed for, analyzing the mean-squared error in the instantaneous predictions and in the turbulent statistics as a function of the distance between the input and target velocity plane. This parametric study is performed at a lower $Re_{\tau}$ of $180$, however a selected case is considered at a higher Reynolds number, where self-similarity can be exploited by the prediction model.

The assumption of self-similarity of the two wall-normal planes at $y/h = 0.2$ and $y/h = 0.1$ (where the length scales are proportional to $y$) at friction Reynolds number $Re_{\tau} \approx 1000$ was the starting point to design an off-wall boundary condition in the model by Mizuno {\it et al.} (2013). In this work, we perform predictions at a comparatively lower Reynolds number of $Re_{\tau} = 550$, with the similar wall-normal planes in scaled outer units. Defining the inner scaling in terms of $u_{\tau}$ and the viscous length $l^{*} = \nu/u_{\tau}$, where $\nu$ is the fluid kinematic viscosity, such wall-normal planes correspond to $y^{+} = 100$ and $y^{+} = 50$, respectively.

\section{Methodology}
\begin{figure*}[ht]
\begin{center}
\includegraphics*[width=0.92\linewidth]{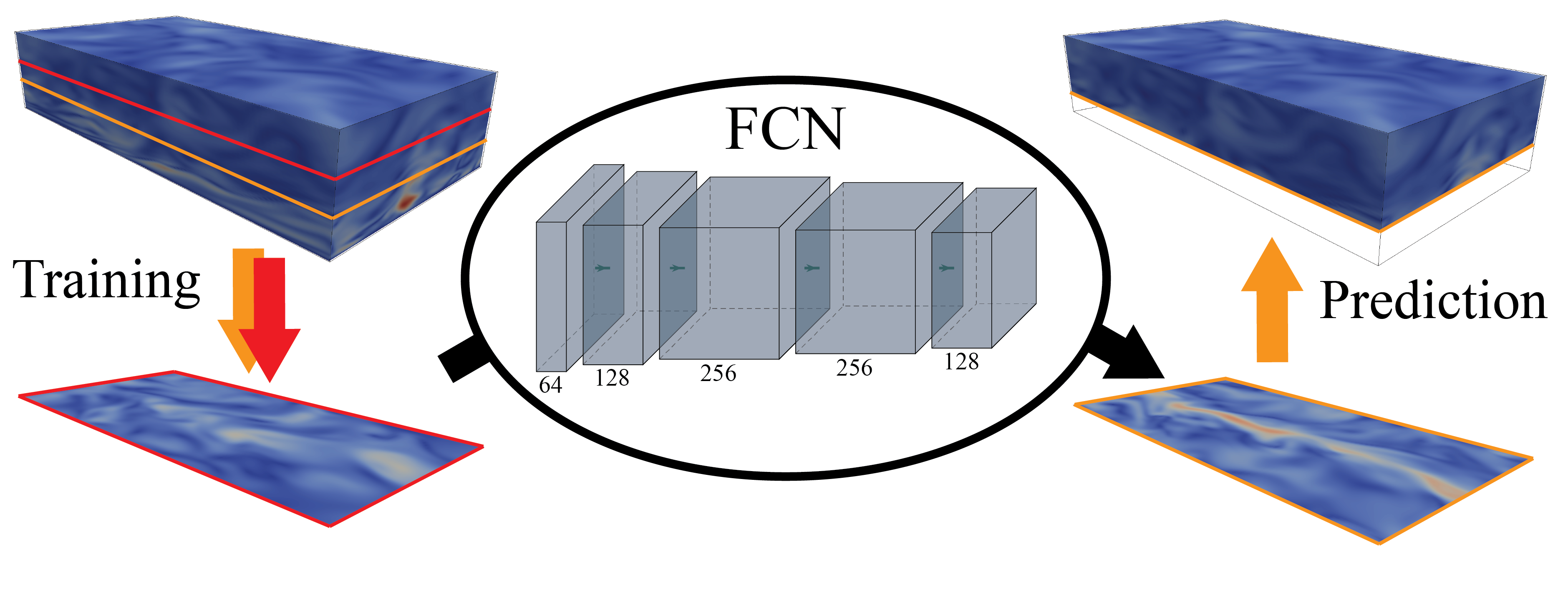}
\caption{\label{cnn_arch} Schematic representation illustrating the use of the FCN architecture. We indicate the number of kernels applied to each of the layers. The kernels (not represented in the Figure) have size $(5\times 5)$ in the first convolutional layer, and $(3\times 3)$ in the subsequent layers.}   
\end{center}
\end{figure*}

The FCN proposed by Guastoni {\it et al.} (2020{\it b}) is shown in figure~\ref{cnn_arch}. This architecture adopts convolution operations in each layer (LeCun {\it et al.} 1998). Convolutions are defined by their kernels (or filters), which contain the learnable parameters, and the transformed output is called \textit{feature map}. Multiple feature maps are usually stacked and followed by an element-wise activation function to make each layer a non-linear transformation. The stacked feature maps are sequentially fed into another convolutional layer as input, which allows the next layer to combine the features individually identified in each map, thus enabling the prediction of larger and more complex features for progressively deeper networks. A batch normalization is performed after each convolutional layer (except for the last one).

The FCN network is trained to predict two-dimensional velocity-fluctuation fields at the given wall-normal location $y^{+}_{\rm target}$, using velocity-fluctuation fields farther from the wall, at $y^{+}_{\rm input}$, as inputs ($y^{+}_{\rm input} > y^{+}_{\rm target}$).
The velocity-fluctuations fields are sampled from a direct numerical simulation (DNS) of an open channel flow, at Reynolds numbers $Re_{\tau} = 180$ and $Re_{\tau} = 550$, performed with the pseudo-spectral solver SIMSON (Chevalier {\it et al.} 2007). The reference length $h$ for this flow case is the open-channel height.
The fields are sampled at wall-normal locations, $y^{+}=15,30,50,80,100,120$ and $150$ at $Re_{\tau} = 180$. For $Re_{\tau} = 550$, only the fields at $y^{+} = 50, 100$ are considered. The resolution of the fields is $(N_{x},N_{z}) = (192,192)$ for $Re_{\tau} = 180$ and $(512, 512)$ for $Re_{\tau} = 550$.
The velocity fields are stored with a constant sampling period of $\Delta t_{s}^{+} = 5.08$ for $Re_{\tau} = 180$ and at $\Delta t_{s}^{+} = 1.49$ for $Re_{\tau} = 550$. The network was trained using 25,200 instantaneous fields for $Re_{\tau} = 180$ and with 19,920 fields for $Re_{\tau} = 550$, split into training and validation sets, with a ratio of 4 to 1. The mean-squared error (MSE) between the instantaneous DNS fields $(u_{\mathrm{DNS}})$ and the predictions $(u_{\mathrm{FCN}})$ is used as the loss function for training the FCN network. It can be written as:
\begin{equation}
\begin{aligned}
\mathcal{L}(&u_{\mathrm{FCN}};u_{\mathrm{DNS}}) = \\
&\frac{1}{N_{x}N_{z}} \sum_{i=1}^{N_{x}} \sum_{j=1}^{N_{z}} | u_{\mathrm{FCN}}(i,j) - u_{\mathrm{DNS}}(i,j) |^{2}\,.
\end{aligned}
\end{equation}

The performance of the network is evaluated based on the quality of the instantaneous predictions in the test dataset, the samples of which are obtained from the simulations initialized with different random seeds ensuring that the training and test dataset do not exhibit unwanted correlations. A total of 1,600 samples and 3,320 samples are used to obtain the converged statistics for $Re_{\tau} = 180$ and $Re_{\tau} = 550$, respectively. The sampling period for the test dataset is the same as that used for the training. It should be noted that the quality of instantaneous predictions improves when the network is trained with less-correlated samples (\textit{i.e. higher $\Delta t_{s}^{+}$}), provided that the network has sufficient capacity (Guastoni {\it et al.} 2020{\it a}). The predictions are also evaluated from the statistical point of view, namely considering the error in RMS quantities of velocity fluctuations:
\begin{equation}
\begin{aligned}
E_{\mathrm{RMS}} (u) = \frac{|u_{\mathrm{RMS,FCN}}-u_{\mathrm{RMS,DNS}}|}{u_{\mathrm{RMS,DNS}}}\,,
\end{aligned}
\end{equation}
and the instantaneous correlation coefficient between the predicted and the DNS fields:
\begin{equation}
\begin{aligned}
R_{\mathrm{FCN; DNS}} (u) = \frac{\left<u_{\mathrm{FCN}} u_{\mathrm{DNS}}\right>_{x,z,t}}{u_{\mathrm{RMS,FCN}} u_{\mathrm{RMS,DNS}}}\,,
\end{aligned}
\end{equation}
with $\left<\cdot\right>$ corresponding to the average in space or time, depending on the subscript. Finally, the pre-multiplied energy spectra of the velocity components are computed for the $Re_{\tau} = 550$ case, in order to verify how the different length scales are predicted by the FCN.

Since the velocity profile does not exhibit a clear logarithmic region, the self-similarity hypothesis used by Mizuno {\it et al.} (2013) is not readily applicable at $Re_{\tau} = 180$, however dataset generation and training are computationally cheaper and the case can be used to check the network implementation.

\section{Results}
In order to understand how the MSE is affected by the separation distance, a total of 21 predictions are performed with the velocity fields obtained at various wall-normal locations from the DNS simulation and we refer to them as \textit{inner} predictions, with different separation distances defined as: $\Delta y^{+} = y^{+}_{\rm input} - y^{+}_{\rm target}$. The input velocity-fluctuation field closest to the wall is located at $y^{+} = 30$ and the corresponding target field at $y^{+} = 15$. The training of the FCN used for inner predictions consisted of 30 epochs, where an epoch identifies a complete pass through the data in the training dataset.

\begin{figure}[htb!]
    \includegraphics[width=0.495\textwidth]{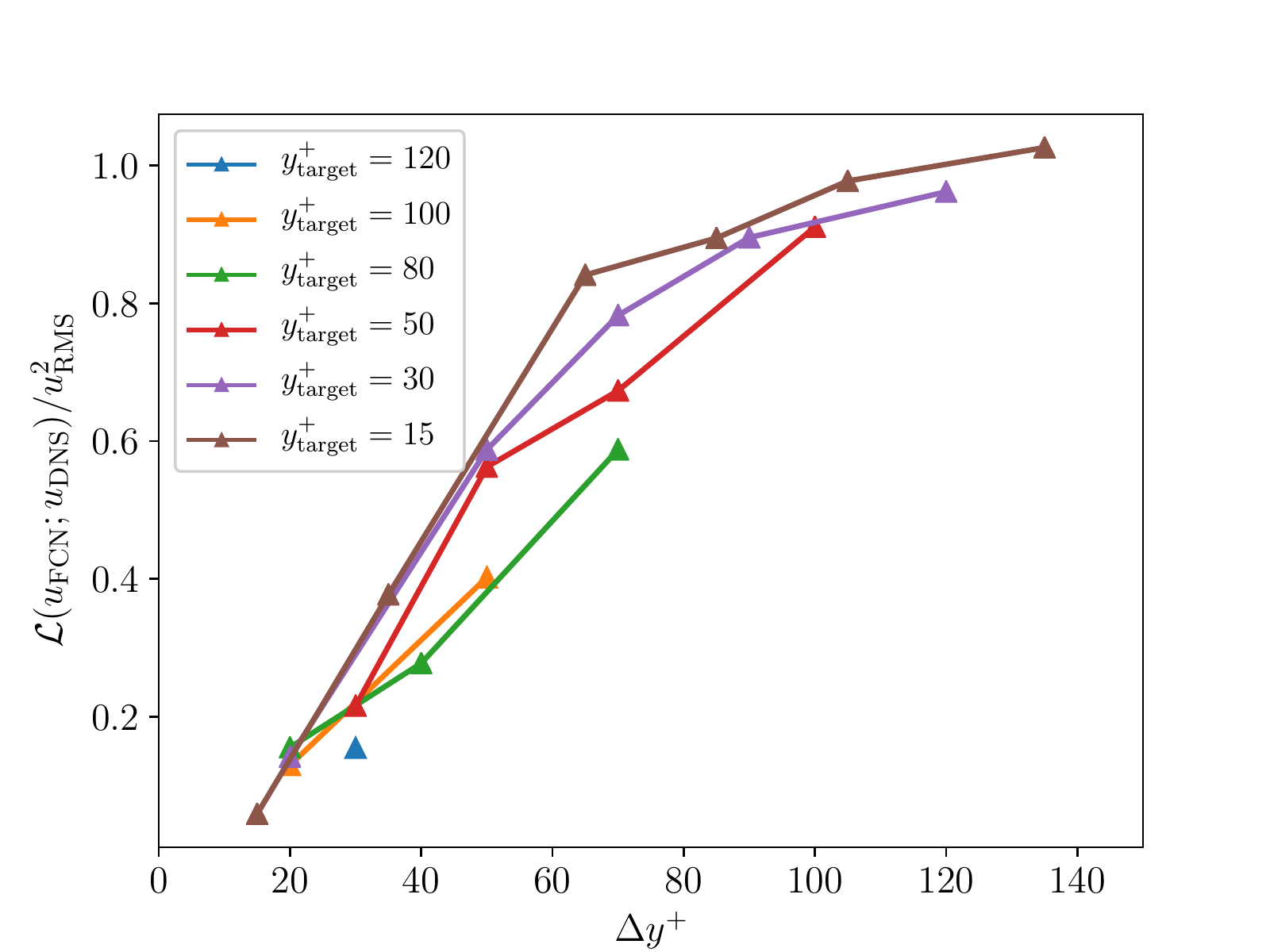}\\
    \includegraphics[width=0.495\textwidth]{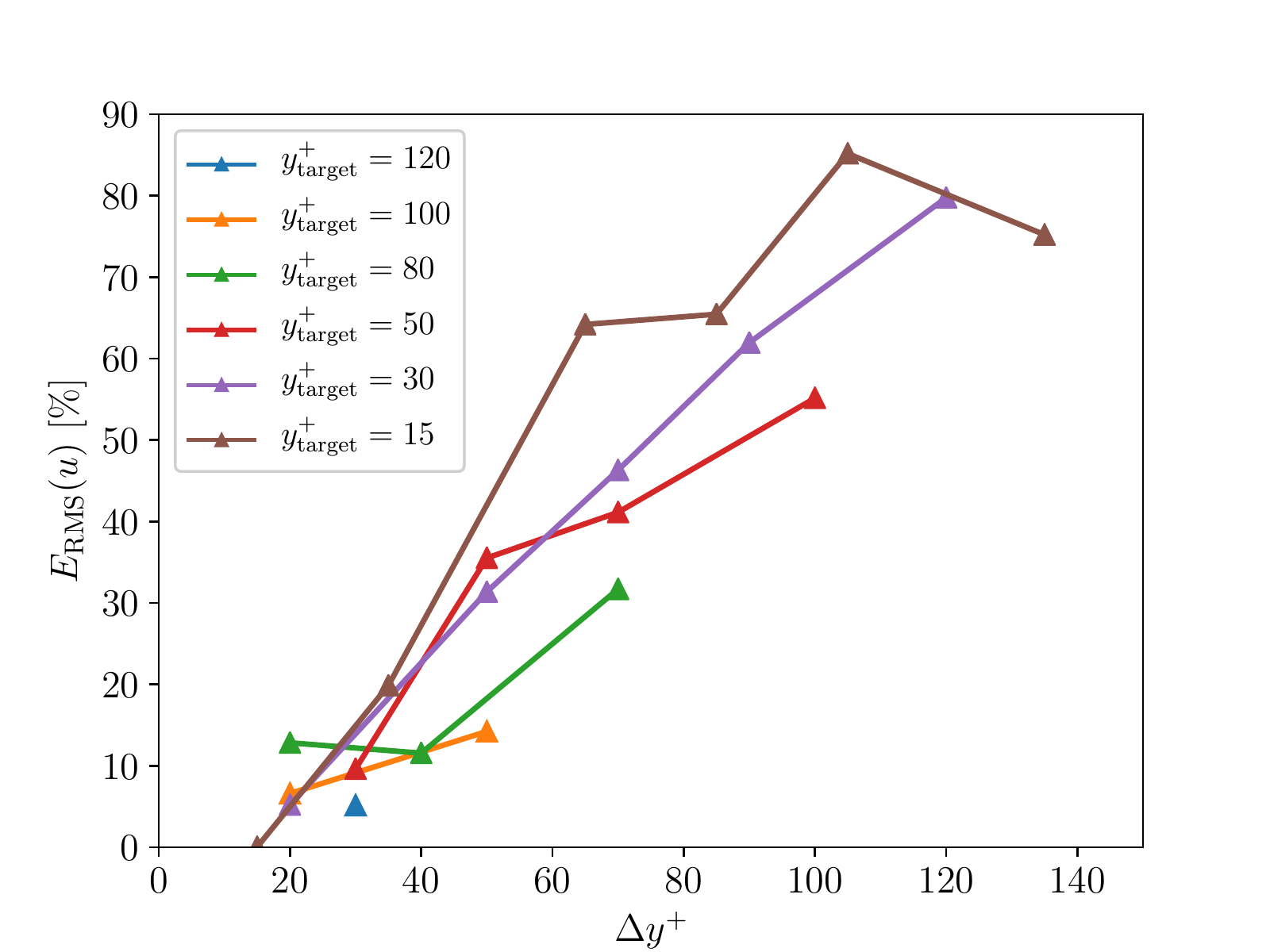}\\
    \includegraphics[width=0.495\textwidth]{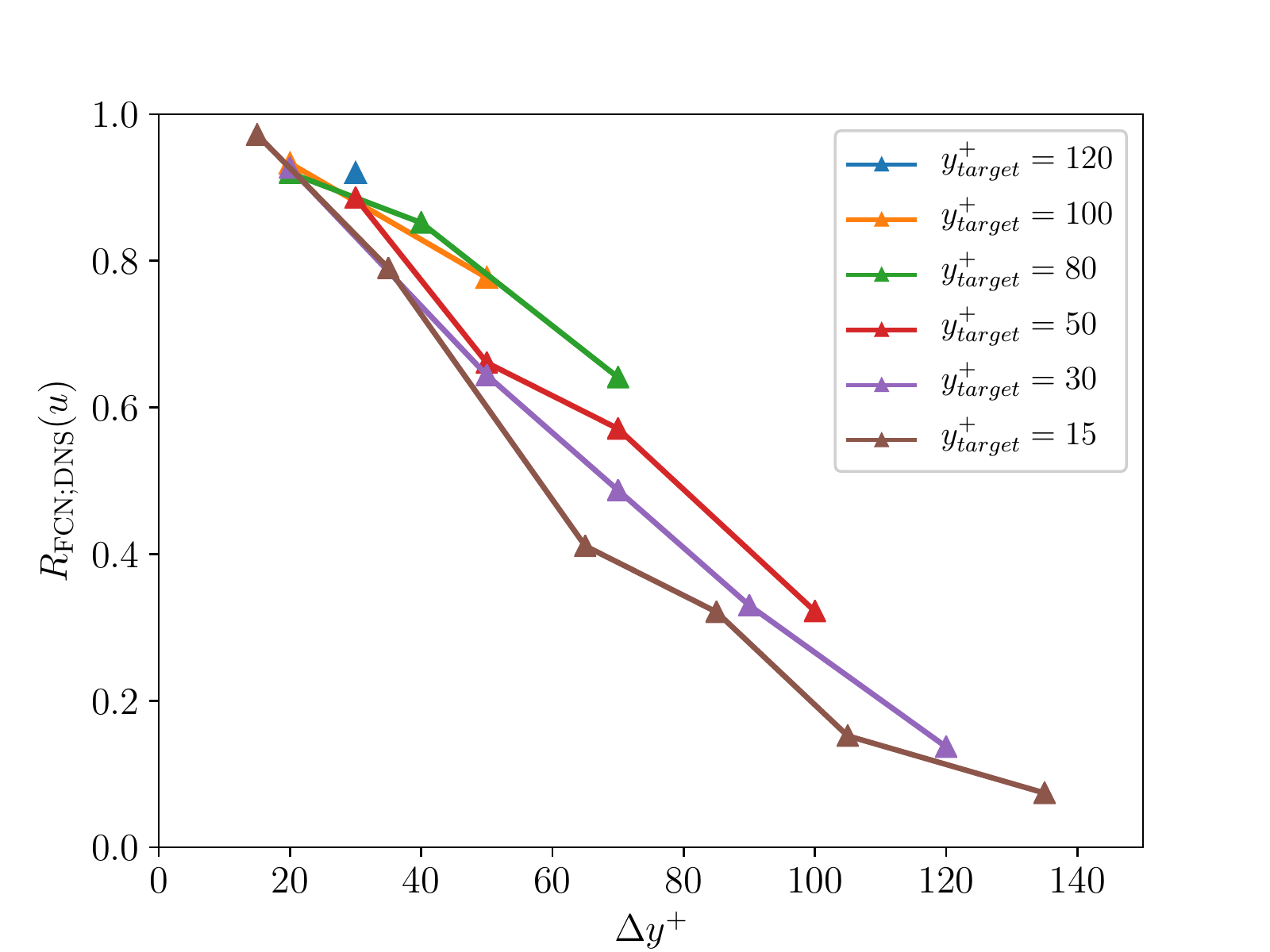}\\
    \caption{\label{loss1}Variation of (top) mean-squared error normalized with the square of the RMS, (middle) relative error in prediction of RMS fluctuation, (bottom) correlation coefficient between the predicted and DNS fields for streamwise velocity component with respect to separation distance for inner predictions.}
\end{figure}

The mean-squared error in the instantaneous predictions, the relative percentage error in the prediction of root-mean-squared (RMS) fluctuations and the correlation coefficient between the predicted fields from FCN and the DNS fields are shown in figure~\ref{loss1}, for the streamwise component of velocity fluctuations. The MSE grows almost linearly for smaller values of $\Delta y^{+}$, whereas for larger values the slope is reduced. From a qualitative analysis of the predictions, the performance is significantly degraded for $\Delta y^{+} > 50$. 
This is perhaps more evident when the statistical error is considered. For larger separation distances, the error is above 30\% and in particular, the intensity of the fluctuations ($u_{\mathrm{RMS,FCN}}$) is severely under-predicted compared to the DNS data ($u_{\mathrm{RMS,DNS}}$). One exception to this trend is the prediction of the fields at $y^{+} = 100$ using information at $y^{+} = 150$. However, these two planes are both located in the wake region, where the variation of the flow features is less pronounced than closer to the wall, for a similar $\Delta y^{+}$.
The correlation coefficient between the predicted and the DNS fields shows a trend similar to that of the MSE. For smaller separation distances, the correlation coefficient is high and close to 1, then it rapidly decreases with $\Delta y^{+}$. From the observation of figure~\ref{loss1}, we can conclude that a correlation coefficient $R_{\mathrm{FCN; DNS}}(u) \approx 0.8$ is the minimum requirement to have convincing predictions. Out of the three velocity components predicted by the FCN network, the streamwise result is used as a reference because it is the most energetic one, influencing most of the overall network performance. 
The other two velocity components predicted along with the streamwise one are typically worse in terms of both statistical error and correlation coefficient.

When the input and the target fields are separated by a large wall-normal distance ($\Delta y^{+} > 50$), it can be difficult for the FCN to learn the relation between input and output. One possible explanation for this is the poor correlation between the input and the target fields for larger separation distances. For smaller $\Delta y^{+}$, there is a significant correlation between the flow features while for larger $\Delta y^{+}$ the correlation drops drastically and the quality of predictions is reduced. The study by Sasaki {\it et al.} (2019) showed the lack of coherence (analogous to correlation and defined in frequency domain) in the short wavelengths (\textit{i.e.} in the small scales) between the input and target velocity fields for large $\Delta y^{+}$. These observations indicate that for larger $\Delta y^{+}$, it would be harder to predict the fields using FCN. Additionally, the higher errors in the prediction of the target fields with larger separation distances can be attributed to the limitation of the convolutional operation when it is used to predict the smaller scales in the target plane (which is closer to the wall). This is because the convolutional operation acts as a filter on the input data. Hence, it becomes difficult for the network to predict the high-frequency content in the output.

If we consider the results at a given separation distance, the MSE in the streamwise component of velocity fluctuation increases as the predicted velocity field is closer to the wall. This is because the velocity-fluctuation fields closer to the wall are characterized by smaller scales and it is harder to predict them from the input velocity fields that may not clearly exhibit such behaviour. Additional complexity arises due to the reduced variability of features in the receptive field (the region of input field from which a single point in the target is obtained) of the input velocity planes, compared to the target velocity planes closer to the wall. We verified this hypothesis by inverting the input and target velocity fields (so that $y^{+}_{\rm input} < y^{+}_{\rm target}$) and training the FCN model to predict the flow farther away from the wall, using near-wall velocity fields as inputs. These predictions are named \textit{outer} predictions and they are similar to the ones performed by (Guastoni {\it et al.} 2020{\it b}) except that the inputs are velocity-fluctuation fields at a given wall-normal distance, instead of the wall-shear stresses and the wall pressure. The inner predictions are more complicated because the convolutional layers filter the content of the input fields and the smaller scales have to be inferred from the larger ones. Indeed, FCN provides a lower MSE in outer predictions compared to inner predictions due to the wide range of small scales in the input velocity fields closer to the wall.

\begin{figure*}[ht]
\begin{center}
\includegraphics*[width=1.0\linewidth]{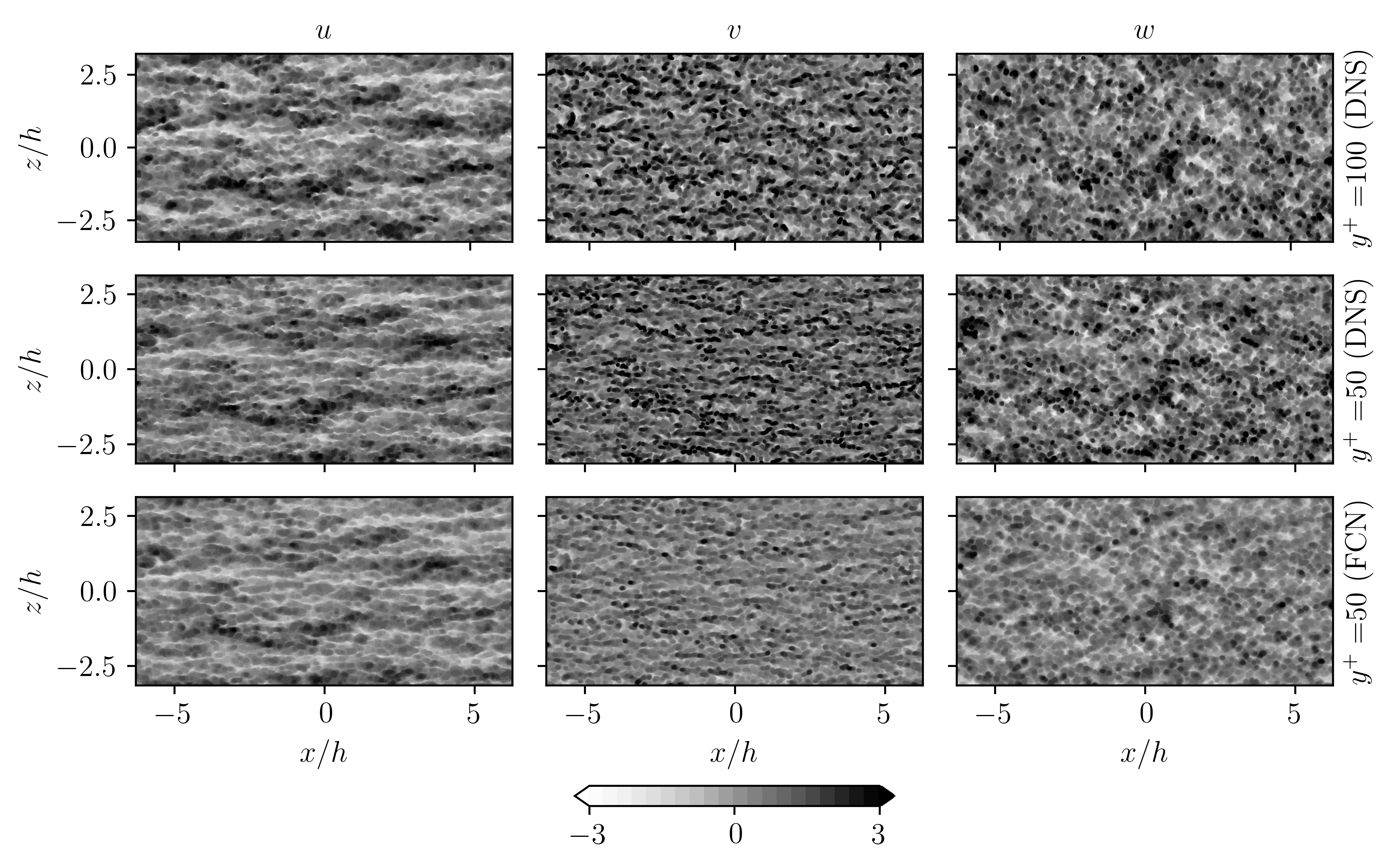}
\caption{\label{fcn_550} (Top row) DNS velocity fluctuations at $y^{+} = 100$, (middle row) at $y^{+} = 50$ and (bottom row) corresponding FCN predictions of the (left column) streamwise, (middle column) wall-normal and (right column) spanwise velocity fluctuations at $y^{+}=50$ for $Re_{\tau} = 550$. The fields are scaled with respect to the corresponding RMS values.}   
\end{center}
\end{figure*}

From the previous results, we find that it is challenging for the FCN to find a non-linear transfer function between the input and output fields with larger $\Delta y^{+}$. At higher Reynolds number, despite the larger range of scales in the flow field, this task becomes simpler because it is possible to exploit the linear dependence of eddy size with respect to the distance from the wall within the logarithmic layer. To this end, we consider the predictions at $y^{+}=50$ $(y/h = 0.1)$ using the velocity-fluctuation fields at $y^{+} = 100$ $(y/h = 0.2)$ for $Re_{\tau} = 550$. These wall-normal locations are similar to the boundary and the reference planes considered by Mizuno {\it et al.} (2013) for the studied $Re_{\tau}$. It should be noted that the linear dependence was originally hypothesized at an asymptotic limit, at very high Reynolds number but, it was also used for finite Reynolds numbers. In figure~\ref{fcn_550} we show a qualitative comparison of the streamwise, wall-normal and spanwise velocity fluctuations predicted by the FCN, compared with the reference DNS data. The quality of prediction is quantitatively assessed using the performance metrics introduced above and they are summarized in table~\ref{ret550_result} for the predictions at $y^{+} = 50$. Since the training procedure is stochastic, the reported errors in the predictions are averaged over 3 different models obtained with different initializations of the learnable parameters of the network.

\begin{table*}[hbt!]
\begin{center}
\caption{\label{ret550_result} Model-averaged errors in the prediction of $y^{+} = 50$ from $y^{+} = 100$ at $Re_{\tau} = 550$}
\begin{tabular}{cccc}
& & \multicolumn{1}{c}{${i}$} & \\ \cline{2-4}
Parameters           & ${u}$                    & ${v}$                      &${w}$\\
\hline
$\mathcal{L}({i}_\mathrm{FCN}; {i}_\mathrm{DNS})/i_\mathrm{RMS}^2$ & 0.365 $\pm$ 0.005 & 0.548 $\pm$ 0.005 & 0.468 $\pm$ 0.006 \\
$E_{\mathrm{RMS}} ({i}) \hspace{0.1cm} [\%] $ & 19.04 $\pm$ 1.17 & 31.8 $\pm$ 1.25 & 25.7 $\pm$ 1.18 \\
$R_{\mathrm{FCN;DNS}}$ & 0.801 $\pm$ 0.003 & 0.681 $\pm$ 0.004 & 0.733 $\pm$ 0.004 \\
\end{tabular}
\end{center}
\end{table*}

\begin{figure}[htb!]
    \includegraphics[width=0.435\textwidth]{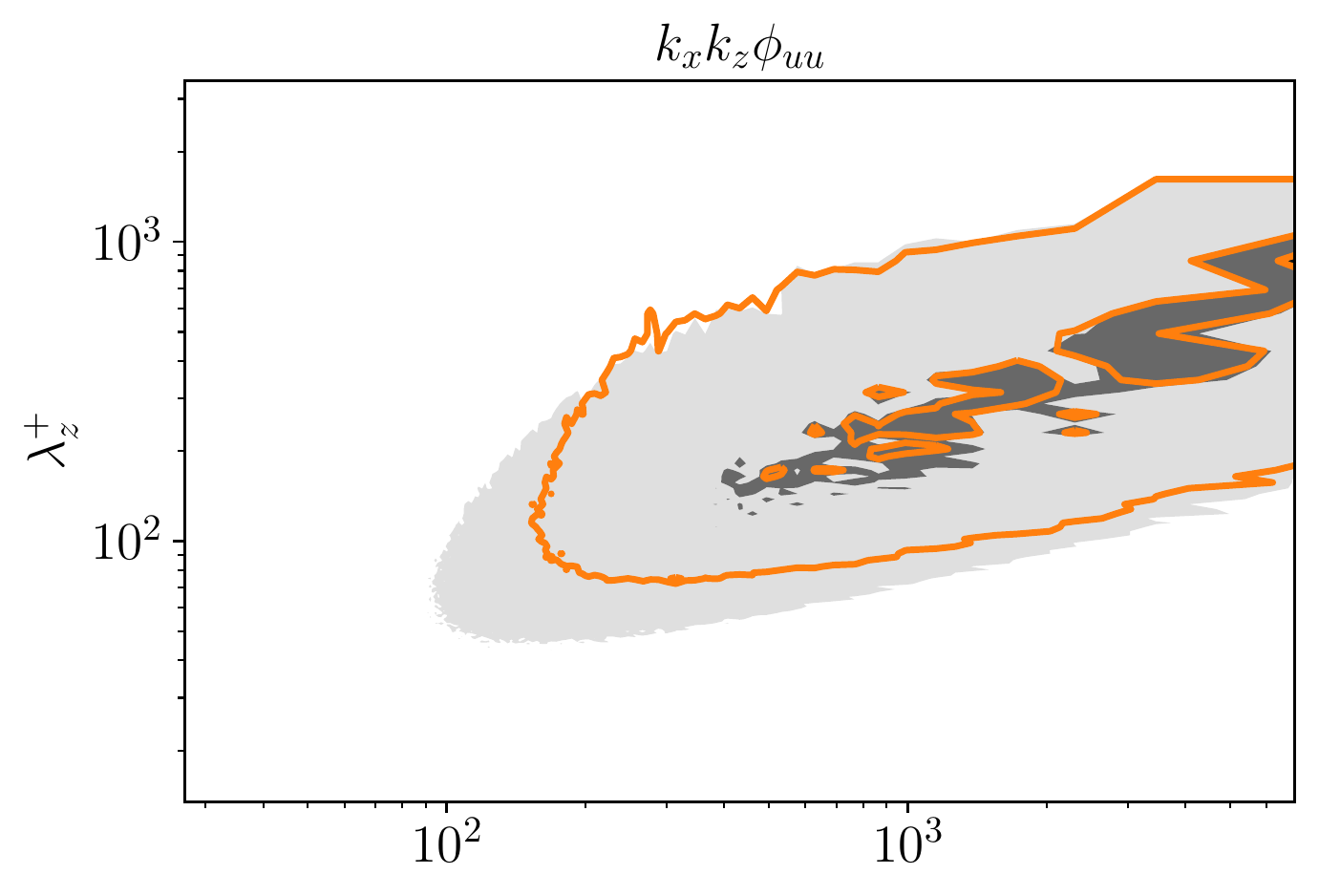}\\
    \includegraphics[width=0.435\textwidth]{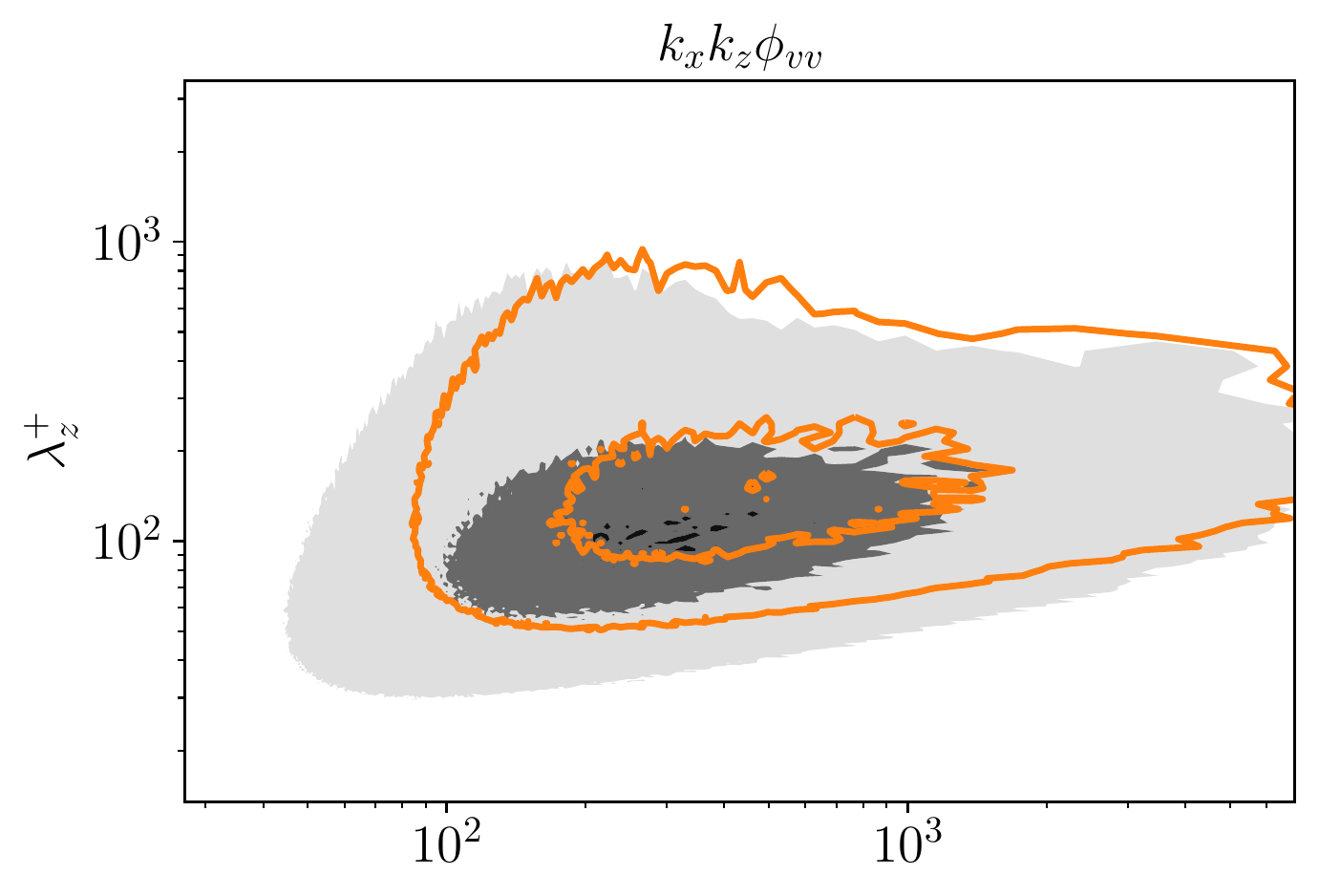}\\
    \includegraphics[width=0.435\textwidth]{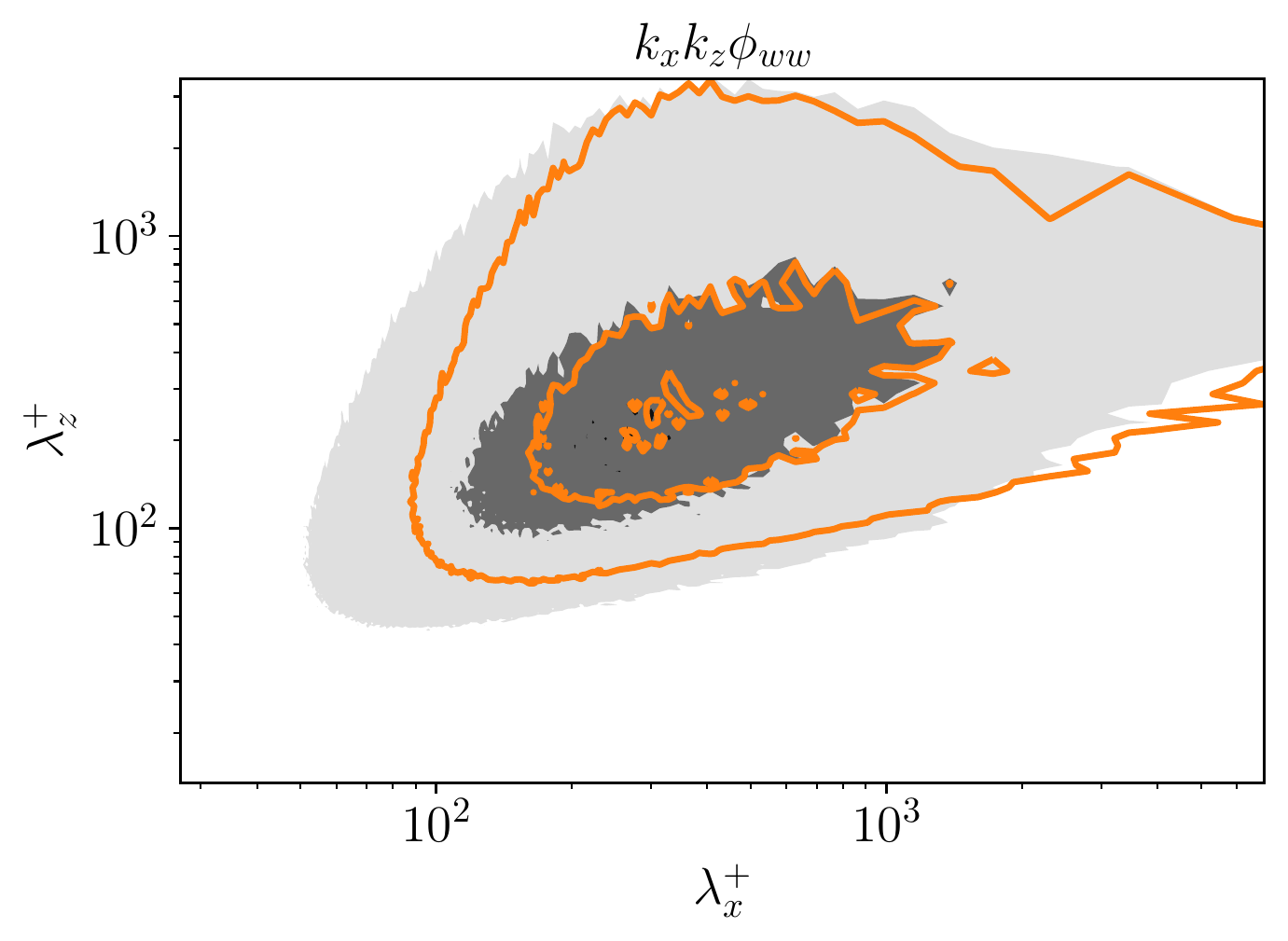}\\
    \caption{\label{spectra} Pre-multiplied two-dimensional power-spectral densities for $Re_{\tau} = 550$ at $y^{+} = 50$. The contour levels contain 10\%, 50\% and 90\% of the maximum DNS power-spectral density. Shaded contours refer to DNS data and dashed contour lines correspond to FCN predictions.}
\end{figure}

At $y^{+} = 50$, the relative error in the RMS streamwise velocity fluctuations from the FCN is less than 20\%, much lower than the 35\% error in the low~$Re$ case. Also, the correlation between the predicted and the DNS velocity-fluctuation field in the streamwise direction is more than 80\% as can also be observed in table~\ref{ret550_result}. Note that the FCN is not explicitly optimized to reproduce the statistical behaviour of the flow, so the results in this metric depend entirely on the capability of the neural network to predict the instantaneous velocity-fluctuation fields. From figure~\ref{fcn_550}, it can be observed that the large scales in the streamwise velocity-fluctuation fields are well represented in the predictions from FCN. The predictions in the smaller scales are also in good agreement with the DNS reference. On the other hand, this is not true for the predictions of the wall-normal and the spanwise velocity-fluctuation fields, which appear smoother than their DNS counterpart, as if a low-pass filter was used. This qualitative analysis is confirmed by the pre-multiplied two-dimensional power-spectral density of the streamwise, wall-normal and spanwise fluctuations shown in figure~\ref{spectra}. From the spectra, it is possible to observe how the energy content of the velocity fields is better reproduced in the streamwise direction, although all three velocity-component predictions lack energy at the shortest wavelengths.

This result highlights that the self-similarity can be implicitly utilized by the FCN in predicting the velocity-fluctuation fields in the overlap region at higher $Re_{\tau}$. 
These preliminary results indicate the advantage in incorporating the physical knowledge available for wall-bounded flows during the development of prospective data-driven wall models for LES. 

\section{Conclusions}
In this work, the possibility of predicting the velocity-fluctuation fields closer to the wall using the fluctuation fields farther from the wall by means of an FCN is assessed. Several predictions were performed to understand the implementation and the quality of predictions at lower $Re_{\tau}$. The variation of the MSE with respect to the separation distance between the input and the target fluctuation fields shows a non-linear behaviour and it also varies with respect to the $y^{+}$ location of the predicted fluctuation field. The results also indicate the capability of the FCN to predict the non-linear transfer function between the velocity-fluctuation fields that are separated by short distances. For good predictions of velocity fields separated by large wall-normal distances, auxiliary losses at intermediate wall-normal distances can be helpful and further investigation in this direction is required. However, additional data requirement would be a limitation in such an approach.

Additionally, a higher $Re_{\tau}$ of 550 was also considered for the same predictions. At high Reynolds number, the self-similarity hypothesis in the logarithmic layer can be exploited by the FCN. The neural network is used to predict the velocity-fluctuation fields at $y^{+}=50$ using the fluctuation fields at $y^{+}=100$ and it provides better instantaneous predictions than the lower-Reynolds case. The fluctuation intensities are also reasonably well predicted, where the error in the streamwise RMS fluctuation is less than 20\% compared to the DNS statistics. Furthermore, the spectral analysis of the predictions shows that the energy content at the different scales is well predicted, with the exclusion of the smaller scales, which are not present in the input flow field.

The present results highlight that the self-similarity in the overlap region of the flow can be effectively utilized by the FCN in predicting the velocity-fluctuation fields at higher $Re_{\tau}$. Note that such similarity is not explicitly enforced in the neural-network architecture. While these preliminary results indicate the potential of FCN as a computationally-effective tool to predict turbulent velocity fields closer to the wall, other architectures might be better suited for the task. As an example, the recent explorations of generational adversarial networks (GANs) applied to super-resolution reconstruction of wall-parallel turbulent fields using coarse data by G\"uemes {\it et al.} (2021), encourage the use of this architecture also in the development of data-driven wall models for LES, since the network is designed to infer smaller features from larger ones.

\Acknowledgments
The authors also acknowledge the funding provided by the Swedish e-Science Research Centre (SeRC), the G\"oran Gustafsson Foundation and the Knut and Alice Wallenberg (KAW) Foundation. SD and AI have been supported by the project ARTURO, ref. PID2019-109717RB-I00/AEI/10.13039/501100011033, funded by the Spanish State Research Agency. The analysis was performed on resources provided by the Swedish National Infrastructure for Computing (SNIC) at PDC. 

\begin{References}
\item Beck, A., Flad, D., and Munz, C. D. (2019). Deep neural networks for data-driven LES closure models. {\it J. Comput. Phys.}, 398 108910.
\item Chevalier, M., Schlatter, P., Lundbladh, A. and Henningson, D. S. (2007), A pseudospectral solver for incompressible boundary layer flows, {\it Tech. rep. TRITA-MEK} 2007:07. KTH Mechanics, Stockholm, Sweden.
\item Guastoni, L., Encinar, M. P., Schlatter, P., Azizpour, H., and Vinuesa, R. (2020{\it a}). Prediction of wall-bounded turbulence from wall quantities using convolutional neural networks. {\it In J. Phys.: Conf. Ser.} 1522, 012022.
\item Guastoni, L., G\"uemes, A., Ianiro, A., Discetti, S., Schlatter, P., Azizpour, H., and Vinuesa, R. (2020{\it b}). Convolutional-network models to predict wall-bounded turbulence from wall quantities, {\it Preprint arXiv:2006.12483}.
\item G\"uemes, A., Discetti, S., Ianiro, A., Sirmacek, B., Azizpour, H., and Vinuesa, R. (2021). From coarse wall measurements to turbulent velocity fields through deep learning. Phys. Fluids, To Appear.
\item Larsson, J., Kawai, S., Bodart, J., and Bermejo-Moreno, I. (2016). Large eddy simulation with modeled wall-stress: recent progress and future directions. {\it Mech. Eng. Rev.}, 3(1), 15-00418.
\item LeCun, Y., Bottou, L., Bengio, Y. and Haffner, P. (1998), Gradient-based learning applied to document recognition, {\it Proc. IEEE} Vol. 86, 2278--2324.
\item Moriya, N., Fukami, K., Nabae, Y., Morimoto, M., Nakamura, T., and Fukagata, K. (2021). Inserting machine-learned virtual wall velocity for large-eddy simulation of turbulent channel flows, {\it  Preprint arXiv:2106.09271}.
\item Mizuno, Y. and Jim\'enez, J. (2013) Wall turbulence without walls, {\it J. Fluid Mech.} Vol. 723, 429--455.
\item Piomelli, U., and Balaras, E. (2002). Wall-layer models for large-eddy simulations. {\it Annu. Rev. Fluid Mech.}, 34(1), 349--374.
\item Sasaki, K., Vinuesa, R., Cavalieri, A.V., Schlatter, P. and Henningson, D.S. (2019). Transfer functions for flow predictions in wall-bounded turbulence, {\it J. Fluid Mech.} Vol. 864, 708--745.


\end{References}

\end{document}